\documentclass{article}
\usepackage{amsmath}
\usepackage{tikz}
\usepackage{cite}
\usetikzlibrary{matrix}
\makeindex
\begin{document}
\title{\bf From Basu-Harvey to Nahm equation via 3-Lie bialgebra }
\author { M.Aali-Javanangrouh$^{a}$
\hspace{-2mm}{ \footnote{ e-mail:
aali@azaruniv.edu}}\hspace{1mm},\hspace{1mm} A.
Rezaei-Aghdam $^{a}$ \hspace{-2mm}{ \footnote{Corresponding author. e-mail:rezaei-a@azaruniv.edu }}\hspace{2mm}\\
{\small{\em
$^{a}$Department of Physics, Faculty of science, Azarbaijan Shahid Madani University }}\\
{\small{\em  53714-161, Tabriz, Iran  }}}
\maketitle
\begin{abstract}
Using the concept of 3-Lie bialgebra; we construct the Bagger- Lambert- Gustavson (BLG) model on the Manin triple $\cal D$  of the especial 3-Lie bialgebra $({\cal D},{\cal A}_{\cal G},{\cal A}_{{\cal G}^*}^*)$ which is in correspondence with Manin triple of Lie bialgebra $({\cal D},{\cal G},{\cal G}^*)$.  We have shown that the Nahm equation (with Lie bialgebra ${\cal G}$) can be obtained from the Basu-Harvey equation as a boundary condition of BLG model (with 3-Lie bialgebra ${\cal D}$) and vice versa.
\end{abstract}

\section{Introduction}
Bagger-Lambert-Gustavsson (BLG) theory is one of the exciting works of the three-dimensional  superconformal field theories that is written as an action for multiple M2-branes \cite{Bagger,Gustavsson}. It has  $N=8$ supersymmetries and SO(8) global symmetry \cite{Lambert}. The main difference of this model  with the others is the utilization of 3- Lie algebra inspired by  Basu-Harvey work \cite{Basu}(see also Ref.\cite{sheikh}). BLG theory has two defects, i)  it only describes two M2-branes \cite{Raamsdonk} and ii) there is only one example for it as $A_4$  \cite{Lambert}, which have been removed by decreasing the number of supersymmetry to $N=6$ \cite{Aharony} and considering Lorentzian metric \cite{Ho}, respectively.

Study of BPS equations has a very important role  in M-theory and string theory \cite{2,3,Strominger}. One can consider these equations as generalizations of Basu-Harvey equation \cite{Basu} and determine the configuration of the M2-M5 bound states with half of the supersymmetries in BLG model \cite{Sezgin}. The BPS conditions for the  D2-D4 configuration are described by Nahm equation \cite{10}. M2-M5 bound states in M-theory are completely related to the bound states in string theory for D2-D4 hence the BPS equations in the BLG theory must be closely related to the Nahm equation  \cite{Copland} as the relation between D2-brane to M2-brane \cite{Mukhi}.  It is possible to obtain Nahm equation from Basu-Harvey equation of the Lorentzian BLG theory\cite{chu} but the inverse is impossible. However, it seems that for the special examples, the inverse is possible by employing the 3-Lie bialgebra, the Lie bialgebra and one to one correspondence between them.

Information about BLG theory can be increased by more studies in 3-Lie algebras. In this respect,  identification of 3- Lie bialgebras would help us as follows \cite{GR}. According to  algebraic work in Ref.\cite{GR} without any role for geometry, it is easy to obtain a reciprocating relation, that has  an important  and applicable  role in M-theory.

  An outline of the paper is as follows. In section 2, using the definition of 3-Lie bialgebra  given in Ref.\cite{GR}, we have considered an especial example and shown that there is a correspondence between this 3-Lie bialgebra and Lie bialgebra. In section 3, we have constructed a BLG model on the Manin triple of this especial 3-Lie bialgebra and obtained Basu-Harvey equation from it and shown that the Nahm equation can be obtained from it and vice versa.

\section{\label{sec:3-Lie bialgebra1}3-Lie bialgebra}
By remembering the definition of Lie bialgebra \cite{Drinfeld} (see for a review Refs. \cite{Sch,Takhtajan,Filippov,vaizman}) we give a short review of the definition of 3-Lie bialgebra \cite{GR}.
\\

{\bf Definition}: \cite{GR} 3-Lie algebra $ \cal A $ with the co commutator map  $\delta : { \cal A} \rightarrow
{ \cal A}\otimes{ \cal A} \otimes{ \cal A} $ is a   \emph{3-Lie bialgebra} if:

{\it a} ) $\delta$ is a 1-cocycle of ${\cal A}$ with value in $ \otimes^3 {\cal A} $, i.e:
\begin{eqnarray}
\label{one cocycle}
\delta([T_{a},T_{b},T_{c}])&=&{ad^{(3)}}_{T_{b}\otimes T_{c}}\delta(T_{a})-{ad^{(3)}}_{T_{a}\otimes T_{c}}\delta(T_{b})+{ad^{(3)}}_{T_{a}\otimes T_{b}}\delta(T_{c}),
\end{eqnarray}
where
\begin{eqnarray}
{ad^{(3)}}_{T_{b}\otimes T_{c}}&=&ad_{T_{b}\otimes T_{c}}\otimes 1\otimes 1+1\otimes ad_{T_{b}\otimes T_{c}}\otimes1+1\otimes1\otimes ad_{T_{b}\otimes T_{c}},
\end{eqnarray}
\\
$\{T_a\}$s are  bases of 3-Lie algebra ${\cal A}$ and  $ ad_{T_a \otimes T_b}T_c=[T_a,T_b,T_c] $, \cite{Takhtajan}.
\\
{\it b} ) the dual map $ ^t\delta:\otimes^3 {\cal A}^* \rightarrow  {\cal A}^*$ is a 3-Lie bracket on ${\cal A}^*$ (dual space of ${\cal A}$ ) with
\begin{equation}
 (\tilde{T}^{a}\wedge \tilde{T}^{b} \wedge \tilde{T}^{c},\delta(T_d))=({^t\delta(\tilde{T}^{a} \wedge \tilde{T}^{b}\wedge \tilde{T}^{c}),T_d)=([\tilde{T}^{a},\tilde{T}^{b},\tilde{T}^{c}]},T_d),
 \label{identity}
\end{equation}
such that it satisfies the fundamental identity. In the above relation $\{{\tilde T}^a\}$ and  $(\hspace{3mm} ,\hspace{3mm} )$ are the basis for the dual space ${ \cal A}^*$ and natural pairing between ${\cal A}$ and $ {\cal A}^*$, respectively. In this way ${ \cal A}^*$ constructs a 3-Lie algebra. The 3-Lie bialgebra can be shown either by $({\cal A} ,{\cal A}^*)$ or $({\cal A},\delta )$.
 \\

{\bf Definition}: \cite{GR} The  Manin triple $({\cal D},{\cal A},{\cal A}^*)$ is  a triple of 3-Lie algebras $({\cal D},{\cal A},{\cal A}^*)$ so that there is a  nondegenerate, symmetric and ad-invariant metric on ${\cal D}$ (Drinfeld double algebra) with the properties\footnote{
Note that in general, the  vector space $\cal D$  is not a 3-Lie algebra \cite{GR}.}:
\\
a) ${ \cal A}$ and ${ \cal A}^*$ are 3-Lie subalgebras of $ {\cal D} $,
\\
b) ${\cal D}={ \cal A}\oplus { \cal A}^*$ as a vector space,
\\
c) ${ \cal A}$ and ${ \cal A}^*$ are isotropic, i.e. \begin{eqnarray}
\nonumber
(T_a,{\tilde T}^b)=\delta_a^b,\hspace{1cm}(T_a,T_b)=({\tilde T}^a,{\tilde T}^b)=0.
\end{eqnarray}
 Using relations  (\ref{one cocycle}), as the fundamental identity, Eq. (\ref{identity})  and $ \delta(T_{a})=\tilde{f}^{ bcd}\hspace{0cm}_a T_{b}\otimes T_{c} \otimes T_{d},
  \label{delta} $ the fundamental and mix fundamental identities for the 3- Lie  bialgebra $({\cal A},{\cal A}^*)$  can be obtained in terms of structure constant (of ${\cal A}$ and ${\cal A}^*$) $f_{abc}\hspace{0cm}^d$ and $\tilde{f}^{abc}\hspace{0cm}_d$ as follows \cite{GR}:
\begin{eqnarray}
f_{aef}\hspace{0cm}^gf_{bcdg}&-& f_{bef}\hspace{0cm}^gf_{acdg}+f_{cef}\hspace{0cm}^gf_{abdg}-f_{def}\hspace{0cm}^gf_{abcg}=0,
\\
{\tilde f}^{aef}\hspace{0cm}_g{\tilde f}^{bcdg}&-& {\tilde f}^{bef}\hspace{0cm}_g{\tilde f}^{acdg}+{\tilde f}^{cef}\hspace{0cm}_g{\tilde f}^{abdg}-{\tilde f}^{def}\hspace{0cm}_g{\tilde f}^{abcg}=0,
\\
\nonumber
{f_{abc}}^{g}\:{\tilde{f}^{def}\:}_{g}&=&{f_{gbc}}^{f}\:{\tilde{f}^{deg}\:}_{a}+{f_{gbc}}^{e}\:{\tilde{f}^{dfg}\:}_{a}
-{f_{gbc}}^{d}\:{\tilde{f}^{efg}\:}_{a}-{f_{gac}}^{f}\:{\tilde{f}^{deg}\:}_{b}+{f_{gac}}^{e}\:{\tilde{f}^{dfg}\:}_{b}
\\
&-&{f_{gac}}^{d}\:{\tilde{f}^{efg}\:}_{b}+{f_{gab}}^{f}\:{\tilde{f}^{deg}\:}_{c}-{f_{gab}}^{e}\:{\tilde{f}^{dfg}\:}_{c}+{f_{gab}}^{d}\:{\tilde{f}^{efg}\:}_{c}.
\label{4-4}
\end{eqnarray}

\subsection{\label{An example}An example}
Now, we will consider an especial example of 3-Lie bialgebra which is constructed of a 3-Lie algebra ${\cal A}_{ \cal G}$ in relation to  Lie algebra ${\cal G}$.
The 3-Lie algebra ${\cal A}_{ \cal G}$ (mentioned in \cite{Ho} for the first time) has commutation relations as  follows:
\begin{eqnarray}
\label{example 3-bracket}
[T_{-},T_{a},T_{b}]=0,\hspace{1cm}
[T_{+},T_{i},T_{j}]={f}_{ij}\hspace{0cm}^{k}T_{k},\hspace{1cm}
[T_{i},T_{j},T_{k}]=f_{ijk}T_{-},
\end{eqnarray}
where $\{T_i\}$s  are the basis  of the Lie algebra ${ \cal G}$ ($[T_{i},T_{j}]={f}_{ij}\hspace{0cm}^{k}
T^{k}$ with $i,j,k=1,2,...,dim { \cal G}$) and $f_{ij}\hspace{0cm}^k$ is its structure constant\footnote{Note that the indices of $f_{ij}\hspace{0cm}^k$ are lowered and raised by the ad-invariant metric  $g_{ij}$ of the Lie algebra $\cal G.$}. Furthermore, $T_{-}$ and $T_+$ are new generators and we have $a=+,-,i$.
Now we propose that  there exists a  3-Lie algebra structure on ${\cal A}^*_{\cal G^*}$ with similar  commutation relations:
\begin{eqnarray}
\label{2example 3-bracket}
[\tilde{T}^{-},\tilde{T}^{a},\tilde{T}^{b}]=0,\hspace{1cm}
[\tilde{T}^{+},\tilde{T}^{i},\tilde{T}^{j}]={\tilde{f}}^{ij}\hspace{0cm}_{k}\tilde{T}^{k},\hspace{1cm}
[\tilde{T}^{i},\tilde{T}^{j},\tilde{T}^{k}]=\tilde{f}^{ijk}T^{-},
\end{eqnarray}
such that  ${\cal G}^*$($[\tilde{T}^{i},\tilde{T}^{j}]={\tilde{f}}^{ij}\hspace{0cm}_{k}
\tilde{T}^{k}$ with $i,j,k=1,2,...,dim { \cal G^*}$) is a Lie algebra.
\par
{\bf Proposition: }{\em  $(\cal{A}_{\cal G},{\cal A}^*_{{\cal G}^*})$ is a 3-Lie bialgebra and the structure constants $f^{abc}\hspace{0cm}_d$ and $\tilde{f}_{abc}\hspace{0cm}^d$ (details of which are given in the following)  satisfy the relation (\ref{4-4}) if and only if $({\cal G},{\cal G}^*)$ is Lie bialgebra, i.e. ${ \cal G}^*$ is a dual Lie algebra of the Lie algebra ${ \cal G}$ , i.e., their structure constants $f^{ij}\hspace{0cm}_k$ and $\tilde{f}_{ij}\hspace{0cm}^k$ satisfy the following Jacobi and  mixed Jacobi identities \cite{Drinfeld,Sch}
 \begin{eqnarray}
f^{ij}\hspace{0cm}_kf^{kl}\hspace{0cm}_m-f^{ik}\hspace{0cm}_mf^{jl}\hspace{0cm}_k+f^{jk}\hspace{0cm}_mf^{il}\hspace{0cm}_k=0,
\label{jacobian identity1}
\\
{\tilde f}_{ij}\hspace{0cm}^k{\tilde f}_{kl}\hspace{0cm}^m-{\tilde f}_{ik}\hspace{0cm}^m{\tilde f}_{jl}\hspace{0cm}^k+{\tilde f}_{jk}\hspace{0cm}^m{\tilde f}_{il}\hspace{0cm}^k=0,
\label{jacobian identity2}
\\
-f^{ij}\hspace{0cm}_k \tilde{f}_{lm}\hspace{0cm}^k+f^{ik}\hspace{0cm}_l \tilde{f}_{km}\hspace{0cm}^j-f^{jk}\hspace{0cm}_m \tilde{f}_{lk}\hspace{0cm}^i-f^{jk}\hspace{0cm}_l \tilde{f}_{km}\hspace{0cm}^i+f^{ik}\hspace{0cm}_m \tilde{f}_{lk}\hspace{0cm}^j=0.
\label{mix jacobian identity}
\end{eqnarray}}

{\bf Proof}:
By expanding the indices $g=i,+,-$ in two sides of the relation (\ref{4-4})   and using the following equalities:
\begin{eqnarray}
\label{f}
\nonumber
f^{+ij}\hspace{0cm}_k&=&f^{ij}\hspace{0cm}_k,\hspace{.75cm}f^{ijk}\hspace{0cm}_-=f^{ijk},\hspace{.75cm}f^{-ab}\hspace{0cm}_c=0,\hspace{.75cm}f^{abc}\hspace{0cm}_+=0,
\\
{\tilde f}_{+ij}\hspace{0cm}^k&=&{\tilde f}_{ij}\hspace{0cm}^k,\hspace{.75cm}{\tilde f}_{ijk}\hspace{0cm}^-={\tilde f}_{ijk},\hspace{.75cm}{\tilde f}_{-ab}\hspace{0cm}^c=0,\hspace{.75cm}{\tilde f}_{abc}\hspace{0cm}^+=0,
\end{eqnarray}
and by taking $,a\equiv i,b\equiv j,d\equiv l,e\equiv m,c=f\equiv k$,  we  arrive at (\ref{mix jacobian identity}), in this way we see that $({\cal A}_{\cal G},{\cal A}^{*}_{\cal G^*})$ is a 3-Lie bialgebra if and only if $(\cal G,\cal G^*)$ is a Lie bialgebra, i.e., ${\cal A}^*={\cal A}^*_{{ \cal G}^*}$. One can consider the Manin triple $({\cal D},{\cal A}_{ \cal G},{{ \cal A}^*}_{{\cal G}^*})$  for this example, i.e. for this example $\cal D$ is a 3-Lie algebra.
\section{From Basu-Harvey to Nahm equation }
In the previous section, we constructed  and showed the existence of a one to one correspondence between the special 3-Lie bialgebra and Lie bialgebra.  In other words,  a special example of 3-Lie bialgebra can be constructed by Lie bialgebra. As is well known,  Bagger, Lambert and Gustavsson \cite{ Bagger, Lambert, Gustavsson} have used 3-Lie algebra instead of Lie algebra for describing multiple M2-branes. In our method, we should construct the BLG theory on the Manin triple $\cal D$, i.e.,  we use 3-Lie bialgebra for describing multiple M2-brane.  Note that the symbol $ F^{ABC}\hspace{0cm}_D $ in the Manin triple ${\cal D}$   apply for the structure constants. In this case,   the Manin triple  is  a $(4+2 dim{{\cal G}})$ dimensional 3-Lie algebra with the  $\{T^A\}$ as  basis of it with $A=i$, $T^i=T^i$ , $A= i+ dim{\cal G}+2$, $T^{ i+ dim{\cal G}+2}=\tilde{T}^i$  , $A= -+dim{\cal G}+2$, $T^{ -+dim{\cal G}+2}=T^{\tilde -}$ , $A=++dim{\cal G}+2$, $T^{++dim{\cal G}+2}=T^{\tilde +}$ and  following commutation relations:
\begin{eqnarray}
\nonumber
[T^-,T^A,T^B]&=&0,\hspace{0.5cm}[T^+,T^i,T^j]=f^{ij}\hspace{0cm}_k T^k,\hspace{0.5cm}[T^+,T^i,T^{\tilde{j}}]= f^{ik}\hspace{0cm}_{j}T^{\tilde{k}},\hspace{0.5cm}[T^i,T^j,T^k]=f^{ijk}T^-,
\\
\label{123}
[T^{\tilde{-}},T^A,T^B]&=&0, \hspace{0.5cm}[T^{\tilde{+}},T^{\tilde{i}},T^{\tilde{j}}]=\tilde{f}_{ij}\hspace{0cm}^kT_{\tilde{ k}},\hspace{0.5cm}[T^{\tilde{+}},T^{\tilde{i}},T^j]={\tilde{f}}^{jk}\hspace{0cm}_i T^k,\hspace{0.5cm}[T^{\tilde{i}},T^{\tilde{j}},T^{\tilde{k}}]=\tilde{f}_{ijk}T_{\tilde{-}}.
\end{eqnarray}
As a first step, consider the following supersymmetric transformation \cite{Bagger,Lambert,Gustavsson}:
\begin{eqnarray}
\label{SUSY}
 \delta X^I_A&=& i\bar{\epsilon} \Gamma^I \Psi_A,
 \nonumber
\\
 \delta\Psi_A&=& D_\mu X^I_A \Gamma^\mu \Gamma_I \epsilon -\frac{1}{2}X^I_BX^J_CX^K_D F^{BCD}\hspace{0cm}_A \Gamma_{IJK}\epsilon ,
 \nonumber
\\
 \delta(\hat{A}_\mu)_B^A&=&i\bar{\epsilon} \Gamma_\mu\Gamma_I X^I_C \Psi_D F^{CDA}\hspace{0cm}_B,
\label{Supersymmetric transformation}
\end{eqnarray}
where $(\hat{A}_{\nu})^B_A=F^{CDB}\hspace{0cm}_A A_{\nu CD}$.  $I,J,...=1,2,...,8$  and  $\mu, \nu, ...=0,1,2$  are the world volume coordinate for M2-brane. $\Gamma_I$ s, $X^{I}$  and $\Psi$ are Dirac matrices, 3-Lie algebra valued transverse coordinates of M2-brane and Majoiarana spinor, respectively with 
$
\Gamma^{012}\Psi=-\Psi,
$
and correspondingly $
\Gamma^{012}\epsilon=\epsilon,
$
 as condition for spinor and  supersymmetric parameter. 
$ D_{\mu} $ is  covariant derivative  which is identified as $
D_{\mu}X^{(I)}_A=\partial_{\mu}X^{(I)}_A+F^{BCD}\hspace{0cm}_AA_{\mu CD}X^{(I)}_B$.
The equations of motion  of BLG model  constructed by the 3- Lie bialgebra valued on Manin triple are the obtained conditions  from  a closed property of the algebra of supersymmetric transformation with the following  Lagrangian \cite{Lambert,Gustavsson}:
\begin{eqnarray}
\nonumber
L=&-&\frac{1}{2}D_{\mu}X^{A(I)}D^{\mu}X_{A}^{(I)}+\frac{i}{2}\bar{\psi}^{A}\Gamma^{\mu}D_{\mu}\psi_{A}+\frac{i}{4}F_{ABCD}\bar{\psi}^{B}\Gamma^{IJ}X^{C(I)}X^{D(J)}\psi^{A}
\\
\nonumber
&-&\frac{1}{12}F_{ABCD}{F_{EFG}}^{D}X^{A(I)}X^{B(J)}X^{C(K)}X^{E(I)}X^{F(J)}X^{G(K)}
\\
&+&\frac{1}{2}\epsilon^{\mu\nu\lambda}[F_{ABCD}{A_{\mu}}^{AB}\partial_{\nu}{A_{\lambda}}^{CD}+\frac{2}{3}{F_{AEF}}^{G}F_{BCDG}{A_{\mu}}^{AB}{A_{\nu}}^{CD}{A_{\lambda}}^{EF}].
\label{BLG  action}
\end{eqnarray}
Evaluation of the supersymmetric boundary condition  results in the supercurrent which  has been  obtained previously as follows \cite{Passerini}:
\begin{equation}
J^\mu =- \bar{\epsilon} D_\nu X^I_A \Gamma^\nu \Gamma^I \Gamma^\mu \Psi^A - \frac{1}{6}
\bar{\epsilon} X^I_A X^J_B X^K_C F^{ABCD}\Gamma^{IJK}\Gamma^\mu \Psi_D\, \ .
\end{equation}
Setting to zero normal component of the boundary of the supercurrent  preserves maximum supersymmetric boundary condition \cite{Sezgin}. 
Assuming  the $x^2$ direction as the boundary, leads to vanishing  $J^2$ for remaining  maximal unbroken  supersymmetry and we will have the following relation:
\begin{equation}
0=\left(- \bar{\epsilon}D_\nu X^I_A \Gamma^\nu \Gamma^I \Gamma^2 \Psi^A - \frac{1}{6} \bar{\epsilon} X^I_A X^J_B X^K_C F^{ABCD}\Gamma^{IJK}\Gamma^2 \Psi_d\right) \left. \right|_{\partial \cal M}.
\label{poi}
\end{equation} 
In order to solve this equation, one should  note the presence of  M2-brane  which  breaks the Lorentz invariance of  M-theory from  $SO(1,10)$ to $SO(1,2)\times SO(8)$  and half of supersymmetry \cite{Sezgin}. In this way,  $SO(1,10)$ breaks to
$SO(1,2) \times SO(8) \rightarrow SO(1,1) \times SO(4) \times SO(4)
$
where $SO(1,1)$ is world sheet of string, one of $SO(4)$ as transverse space for M5-brane and the other as  transverse space for both of M2-brane and M5-brane\footnote{In Ref.  \cite{Sezgin} there is another way of breaking Lorentzian symmetry $SO(1,2) \times SO(8) \rightarrow SO(1,1) \times SO(8)
$ where we don't describe it here.}. The scalar fields $X^I$ are decomposed    as $X^V = \{X^3,X^4,X^5,X^6 \}$ corresponding to SO(4) symmetry and $Y^P = \{ X^7, X^8,X^9,X^{10}\}$ corresponding to the other  $SO(4)$ symmetry, subsequently  the equation (\ref{poi}) turns into the following form \cite{Sezgin}:
\begin{eqnarray}
\nonumber
0&=& -\bar{\epsilon}D_{\hat{\nu}} X^V_A \Gamma^{\hat{\nu}} \Gamma^V \tilde{\Psi}^A  \\
\nonumber
&& - \bar{\epsilon}D_{\hat{\nu}}  Y^P_A \Gamma^{\hat{\nu}}\Gamma^P \tilde{\Psi}^A   \\
\nonumber
&&  - \bar{\epsilon}\left( D_2 Y^P_A \Gamma^2 \Gamma^P \delta^{DA} + \frac{1}{6} Y_A^PY_B^QY_C^R F^{ABCD} \Gamma^{PQR}\right) \tilde{\Psi}_D   \\
\nonumber
&& - \bar{\epsilon}\left( D_2X^V_A \Gamma^2 \Gamma^V \delta^{DA} + \frac{1}{6} X_A^VX_B^UX_C^W F^{ABCD} \Gamma^{VUW}\right) \tilde{\Psi}_D  \\
\nonumber
&&  - \bar{\epsilon}\left( \frac{1}{2} X_A^VX_B^UY_C^P F^{ABCD} \Gamma^{VUP}\right) \tilde{\Psi}_D  \\
&&  - \bar{\epsilon}\left( \frac{1}{2} X_A^VY_B^P Y_C^Q F^{ABCD} \Gamma^{APQ}\right) \tilde{\Psi}_D, 
\label{18}
\end{eqnarray}
where  $\Gamma_2 \Psi_A = \tilde{\Psi}_A$ and $\hat{\mu} = 0,1$, also the terms ordered in each line in terms of their Lorentzian structures and   preserving  $SO(1,1) \times SO(4)\times SO(4)$ symmetry vanishes  separately. In order to solve the equations, the type of boundary condition (the Dirichlet or Neumann condition) must be specified.   Assuming  half of the scalars to obey Dirichlet conditions \footnote{ We will not describe here the Neumann boundary condition, for a review see \cite{Sezgin}.} $D_{\hat{\mu}} Y^P = 0$,  $Y^P= 0$ is obtained as   the simplest solution for these conditions and after substituting this solution  in (\ref{18}) we have the following relations:
\begin{eqnarray}
\label{eqA} 0&=& \bar{\epsilon} D_2 Y^P \Gamma^2 \Gamma^P \tilde{\Psi} \ ,\\
\label{eqB} 0&=& \bar{\epsilon} D_{\hat{\nu}}X^V \Gamma^{\hat{\nu}} \Gamma^V \tilde{\Psi} \ ,  \\
\label{eqC} 0&=& \bar{\epsilon} \left(D_2 X_A^V\delta^{AD} \Gamma^2 \Gamma^V + \frac{1}{6}F^{ABCD} X^V_AX^U_BX^W_C \Gamma^{VUW} \right) \tilde{\Psi}_D \ .
\end{eqnarray}
For solving these  equations  suppose
$\frac{1}{2}(1- \Gamma^{013456} ) \tilde{\Psi}={\cal Q}_- \tilde{\Psi}     =0 \
$
or   $(1-\Gamma^{013456} ) \epsilon={\cal Q}_-\epsilon = 0$ on the supersymmetric parameter, then (\ref{eqB}) holds automatically \cite{Sezgin}.  Using 
$ \Gamma^V = \frac{1}{6} \epsilon^{VUWZ} \Gamma^{UWZ}\Gamma^{3456} $  the equation  (\ref{eqC}) takes the following relation:
\begin{eqnarray}
0= \left(D_2X^Z_A\epsilon^{ZVUW}  \delta^{DA} + X_A^VX_B^UX_C^W F^{ABCD} \right)  \bar{\epsilon}\Gamma^{VUW} \tilde{\Psi}_D, 
\end{eqnarray}
which for the  scalar fields  $X^V$ is  the Basu-Harvey type equations: 
\begin{eqnarray}
\label{Basu}
0 = D_2X^V_A + \frac{1}{6} \epsilon^{VUWZ} X^U_B X^W_C X^Z_D  F^{BCD}_{\phantom{ABC}A} , 
\end{eqnarray}
 we expande it as follows:
\begin{eqnarray}
0 &=& D_2X^V_- + \frac{1}{6} \epsilon^{VUWZ} X^U_B X^W_C X^Z_D F^{BCD}_{\phantom{abc}-} , 
\\
0 &=& D_2X^V_{\tilde -} + \frac{1}{6} \epsilon^{VUWZ} X^U_B X^W_C X^Z_D F^{BCD}_{\phantom{abc}\tilde -} , 
\\
0 &=& D_2X^V_i + \frac{1}{2} g_{YM}\epsilon^{VUW}  X^U_j X^W_k  f^{jk}_{\phantom{abc}i} + \frac{1}{2} g_{YM}\epsilon^{VUW}  X^{U\tilde j} X^W_k  {\tilde f}^{k}_{\phantom{abc}ij}, 
 \\
0 &=& D_2X^V_{\tilde i}+ \frac{1}{2} g_{YM}\epsilon^{VUW}  X^{U\tilde j} X^W{\tilde k}  {\tilde f}_{jk}^{\phantom{abc}\tilde i} + \frac{1}{2} g_{YM}\epsilon^{AUW}  X^{U\tilde j} X^W_k f^{k}_{\phantom{abc}{\tilde i}{\tilde j}},
\end{eqnarray}
where we have used  $i, j, k$ for  representation  of bases of Lie bialgebra ${\cal D}_{\cal G}$. Using relation (\ref{123}), one can add up  relations (24)-(27) to obtain the following relation:
\begin{equation}
0 =\partial_2X^V_{\tilde i}+ \frac{1}{2} g_{YM}\epsilon^{VUW}  X^{U\tilde j} X^W{\tilde k}  {\tilde f}_{jk}^{\phantom{abc}\tilde i} + \frac{1}{2} g_{YM}\epsilon^{AUW}  X^{U\tilde j} X^W_k f^{k}_{\phantom{abc}{\tilde i}{\tilde j}}.
\end{equation}
Let us consider the  Nahm equation as a boundary condition for a system of D2-branes ending on D4-brane which is  related  to  the Lie bialgebra representation as follows:
\begin{eqnarray}
\nonumber
\partial_\sigma X^I _A&=&   
\frac{1}{2} \epsilon_{JK}^IX^J_B X^K_C F^{BC}_A, \\
\partial_\sigma X^I _i&=&   
\frac{1}{2} \epsilon_{JK}^IX^J_j X^K_k f^{jk}_i +\frac{1}{2} \epsilon_{JK}^IX^{J\tilde j} X^{K}_k {\tilde f}^{k}_{i\tilde j} 
,
\end{eqnarray} 
 where $F^{AB}\hspace{0cm}_C$ is the structure constant of ${\cal D}_{\cal G}$ (the  Drinfeld double of Lie algebra $\cal G$) \cite{Sch} and  $A,B,C=i,{\tilde i}$. As it has been done by the algebraic structure in this work, it can be done inversely, i.e., by this method, one can obtain the Nahm equation from the Basu-Harvey equation and vice versa.

The Basu-Harvey and Nahm equations can be considered as BPS bounds for M2-brane and D1-string, respectively.  It would be possible to state a reciprocate relation between these equations as BPS bounds identified in 3- Lie bialgebra, i.e., we want to reach from BPS bound for M2-brane to  BPS bound for  D1-string and vice versa. BPS bound is a result of vanishing supersymmetric transformations of gauge and fermion fields that we obtained it for M2-branes ending M5-brane by using BLG Lagrangian and showed that  the result was similar bound for  D1-string ending D3-brane. The BLG Lagrangian (\ref{BLG action}) regardless of the fermion piece can be rewritten as \cite{Lowthesis}:
\begin{eqnarray}
L=-\frac{1}{2}D_{\mu}X^{A(I)}D^{\mu}X_{A}^{(I)}
-\frac{1}{12}F_{ABCD}{F_{EFG}}^{D}X^{A(I)}X^{B(J)}X^{C(K)}X^{E(I)}X^{F(J)}X^{G(K)}
,
\label{BLG  action without fermion}
\end{eqnarray}
and the equivalent indication  for the energy can be obtained as follows:
\begin{eqnarray}
\nonumber
E&=&\frac{1}{2}Tr(\partial_{s}X^{A(I)}\partial^{s}X_{A}^{(I)})
+\frac{1}{12}Tr([X^{(I)},X^{(J)},X^{(K)}],[X^{(I)},X^{(J)},X^{(K)}])
\\
\nonumber
&+&\frac{1}{2}Tr(\partial_{s}X_{A}^{(I)})
-\frac{1}{12}Tr([X^{(I)},X^{(J)},X^{(K)}])^2+\frac{1}{6}\epsilon^{JIKL}Tr(\partial_sX^{(I)},[X^{(J)},X^{(K)},X^{(L)}])
\\
&\geq&\frac{1}{6}\epsilon^{JIKL}Tr(\partial_sX^{(I)},[X^{(J)},X^{(K)},X^{(L)}])
\end{eqnarray}
i.e., in this way the Basu-Harvey equation is obtained as a BPS bound. 
But the above equation can be written as
\begin{eqnarray}
\nonumber
E
&\geq&\frac{1}{6}\epsilon^{JIKL}Tr(\partial_sX^{(I)}_+T^++\partial_sX^{(I)}_-T^-+\partial_sX^{(I)}_iT^i+\partial_sX^{(I)}_{\tilde+}T^{\tilde+}+\partial_sX^{(I)}_{\tilde-}T^{\tilde-}
\\
\nonumber
&&+\partial_sX^{(I)}_{\tilde i}T^{\tilde i},X^{(J)}_+X^{(K)}_iX^{(L)}_j[T^+,T^i,T^j]+X^{(J)}_+X^{(K)}_iX^{(L)}_{\tilde j}[T^+,T^i,T^{\tilde j}]
\\
\nonumber
&&+X^{(J)}_{\tilde+}X^{(K)}_{\tilde i}X^{(L)}_{\tilde j}[T^{\tilde+},T^{\tilde i},T^{\tilde j}]+X^{(J)}_{\tilde+}X^{(K)}_{\tilde i}X^{(L)}_j[T^{\tilde+},T^{\tilde i},T^j])
\\
&\geq&\frac{1}{2g_{YM}^2}\int d\sigma \epsilon_{IJK}\partial_sX^I[X^J,X^K].
\label{BLG  action without fermion}
\end{eqnarray}
Therefor, we have shown that there is a relation between Basu-Harvey equation as  BPS bound obtained from BLG model for multiple membranes and the Nahm equation as BPS bound obtained from Yang-Mills action for multiple Dp-branes.
 \section*{ Conclusions}
Using the concept of 3-Lie bialgebra, studied in arXiv:1604.04475, we have obtained Nahm equation from Basu-Harvey equation by studying the boundary condition of BLG model, the Manin triple $\cal D$ of 3-Lie bialgebra $({\cal D},{\cal A}_{\cal G},{\cal A}_{{\cal G}^*}^*)$. In this manner, it seems that the concept of 3-Lie bialgebra is a good idea. One can consider the BLG model on $\cal D$ 3-Lie algebra and can obtain the $N=(4,4)$ WZW like a model on Lie bialgebra $({\cal G},{\cal G}^*)$ using the correspondence  of 3-Lie bialgebra $({\cal A}_{\cal G},{\cal A}_{{\cal G}^*}^*)$ with Lie bialgebra v and vice versa, i.e. (D2 $\leftrightarrow$ M2). Another problem is the investigation of the $N=6$ BL model where is the superconformal model \cite{Aharony} using 3-Leibniz bialgebra \cite{GR} and studying the boundary condition.
\subsection*{Acknowledgments} 
We would like to thank   M. Akbari-Moghanjoughi for carefully reading the manuscript. This research was supported by a research fund
No. 217D4310 of Azarbaijan Shahid Madani university.


\begin{thebibliography}{99}
\bibitem{Bagger}J. Bagger, N. Lambert, Modeling Multiple M2's, Phys. Rev. D{\bf 75} (2007) 045020 [arXiv:hep-th/0611108].

\bibitem{Gustavsson}A. Gustavsson, Algebraic structures on parallel M2-branes,  Nucl. Phys. B {\bf811} (2009) 66, [arXiv:0709.1260 [hep-th]].
\bibitem{Lambert}J. Bagger and N. Lambert, Gauge Symmetry and Supersymmetry of Multiple M2-Branes, Phys. Rev. D{\bf 77} (2008) 065008, [arXiv:0711.0955 [hep-th]];  J. Bagger, N. Lambert, Comments on Multiple M2-branes,  JHEP {\bf 02} (2008) 105, [arXiv:0712.3738 [hep-th]].
\bibitem{Basu}A. Basu, J. A. Harvey, The M2-M5 Brane System and a Generalized Nahm's Equation, Nucl. Phys. B{\bf713} (2005) 136 [arXiv:hep-th/0412310].
\bibitem{sheikh}M. M. Sheikh-Jabbari, Tiny Graviton Matrix Theory: DLCQ of IIB Plane-Wave String Theory, A Conjecture, JHEP 0409:017,2004, [	arXiv:hep-th/0406214].
\bibitem{Raamsdonk}M. Van Raamsdonk, Comments on the Bagger-Lambert theory and multiple M2-branes,  JHEP {\bf 05} (2008) 105, [	arXiv:0803.3803 [hep-th]].

\bibitem{Aharony}O. Aharony, O. Bergman, D. L. Jafferis, J. Maldacena,N=6 superconformal Chern-Simons-matter theories, M2-branes and their gravity duals,  JHEP {\bf10} (2008) 091, [	arXiv:0806.1218 [hep-th]]; J. Bagger, N. Lambert, Three-Algebras and N=6 Chern-Simons Gauge Theories,  Phys. Rev. D{\bf79} (2009) 025002, [arXiv:0807.0163 [hep-th]].
 \bibitem{Ho}P.-M. Ho, Y. Imamura, Y. Matsuo, M2 to D2 revisited, JHEP {\bf07} (2008) 003, [arXiv:0805.1202 [hep-th]].
 \bibitem{2} S. Terashima, On M5-branes in N=6 Membrane Action, JHEP 0808 (2008) 080 [arXiv:0807.0197 [hep-th]].
\bibitem{3} J. Gomis, D. Rodriguez-Gomez, M. Van Raamsdonk and H. Verlinde, A Massive Study of M2-brane Proposals,  JHEP
0809 (2008) 113 [arXiv:0807.1074 [hep-th]].
\bibitem{Strominger}P. K. Townsend, D-branes from M-branes,  Phys. Lett. B {\bf373} (1996) 68 [	arXiv:hep-th/9512062]; A. Strominger,  Open P-Branes, Phys. Lett. B {\bf383} (1996) 44  [arXiv:hep-th/9512059]; C. S. Chu and E. Sezgin, M-Fivebrane from the Open Supermembrane,  JHEP {\bf9712} (1997) 001 [arXiv:hep-th/9710223]; C. S. Chu, P. S. Howe, E. Sezgin and P. C. West, Open Superbranes,  Phys. Lett. B {\bf429} (1998) 273 [arXiv:hep-th/9803041].
\bibitem{Sezgin}D. S. Berman, M. J.Perry, E. Sezgin, D. C. Thompson, Boundary Conditions for Interacting Membranes,  JHEP {\bf1004}(2009)20, [arXiv:0912.3504 [hep-th]].
\bibitem{10} W. Nahm, Phys. Lett. B{\bf90} (1980) 413; D. E. Diaconescu, D-branes, Monopoles and Nahm Equations,  Nucl. Phys. B 503 (1997) 220 [arXiv:hep-th/9608163].
\bibitem{Copland}C. S Chu, G. S. Sehmbi, Open M2-branes with Flux and Modified Basu-Harvey Equation, J.Phys.A44:135404,2011, [	arXiv:1011.5679 [hep-th]]; K. Sakai, S. Terashima,  Integrability of BPS equations in ABJM theory, JHEP 11 (2013) 002 , [	arXiv:1308.3583 [hep-th]].
\bibitem{Mukhi}S. Mukhi, C. Papageorgakis, M2 to D2, JHEP {\bf05} (2008) 085 [arXiv:0803.3218 [hep-th]]; Y. Pang, T. Wang, From N M2's to N D2's, Phys. Rev. D{\bf78} (2008) 125007, [arXiv:0807.1444 [hep-th]].
\bibitem{chu}C.S. Cho and  D. J Smith, Open M2-branes with Flux and Modified Basu-Harvey Equation,  J.Phys.A44:135404,2011, 	[arXiv:1011.5679 [hep-th]].
\bibitem{GR} A. Rezaei-Aghdam, L. Sedghi-Ghadim,  {\it 3-Leibniz bialgebra (3-Lie bialgebra)}, 	[	arXiv:1604.04475 [math.RA]].
\bibitem{Drinfeld} V. G. Drinfeld, Vol. {\bf1} Proceedings of the International Congress of Mathematicians, Berkeley, (1986) 789.
\bibitem{vaizman}I. Vaisman,  Progress in Mathe- matics, Birkh¨auser
Basel, Vol {\bf118} (1994).
\bibitem{Sch} Y. K. Schwarzbach,  Lecture notes in physics {\bf038}, Springer-Verlag (2004)107.
\bibitem{Takhtajan}L. Takhtajan, Comm. Math. Phys. {\bf160} (1994), 295.
\bibitem{Filippov} V. Filippov, Sibirsk. Mat. Zh. {\bf26} (1985) 126140.

\bibitem{Passerini}F. Passerini, M2-Brane Superalgebra from Bagger-Lambert Theory, JHEP {\bf08} (2008) 062,[	arXiv:0806.0363 [hep-th]].
\bibitem{Lowthesis}A. M. Low,  PH.D thesis, Queen Mary University of London [hep-th/1012.2707].






\end{thebibliography}
\end{document}